\documentclass[11pt, letter]{article}
\usepackage[utf8]{inputenc}
\usepackage[final]{microtype}
\usepackage{authblk}
\usepackage[letterpaper, margin=1in]{geometry}

\usepackage[hyphens]{url}
\usepackage[square,numbers]{natbib}
\usepackage[hidelinks]{hyperref}
\usepackage{breakurl}

\usepackage{booktabs}

\usepackage{graphicx}

\usepackage{url}

\usepackage[inline]{enumitem}

\usepackage{amsmath,amsfonts,amssymb,amsthm}
\DeclareMathOperator{\poly}{poly}

\title{Quantum Computing for Software Engineering: Prospects}
\begin{document}
\author[1]{Andriy Miranskyy}
\author[2]{Mushahid Khan}
\author[3]{Jean Paul Latyr Faye} 
\author[4]{Udson C. Mendes} 

\affil[1]{Department of Computer Science, Toronto Metropolitan University (formerly~Ryerson~University), Toronto, ON M5B2K3 Canada}
\affil[2]{Department of Computer Science, Toronto Metropolitan University (formerly~Ryerson~University), Toronto, ON M5B2K3 Canada}
\affil[3]{CMC Microsystems, Sherbrooke, QC J1K1B8 Canada}
\affil[4]{CMC Microsystems, Sherbrooke, QC J1K1B8 Canada}
\affil[ ]{\textit{avm@ryerson.ca, mushahid.khan@ryerson.ca, jean.paul.latyr.faye@cmc.ca, udson.mendes@cmc.ca}}
\date{}
\maketitle

\begin{abstract}
Quantum computers (QCs) are maturing. When QCs are powerful enough, they may be able to handle problems in chemistry, physics, and finance that are not classically solvable. However, the applicability of quantum algorithms to speed up Software Engineering (SE) tasks has not been explored. 
We examine eight groups of quantum algorithms that may accelerate SE tasks across the different phases of SE and sketch potential opportunities and challenges. 

\end{abstract}

\section{Introduction}\label{sec:intro}

Quantum computing is a nascent but rapidly growing field. The quantum computing market size is expected to reach USD 1.7 billion in 2026, up from USD 0.5 billion in 2021, at a compound annual growth rate of 30.2\%~\cite{markets2021quantum}. 

In order for quantum computers (QCs) to become practical and solve real-world problems, the software running on them must be diverse and of high quality. Thus, an exploration of bringing Software Engineering (SE) practices into the quantum computing community is in order~\cite{miranskyy2019testing, zhao2020quantum, miranskyy2021testing}. These practices will enable QC programmers to write better-quality code. 

But is the inverse possible? That is, can quantum computing algorithms be used to speed up SE tasks? The potential benefits of QCs to, e.g., chemists, physicists, and financiers have been extensively studied~\cite{ORUS2019100028, preskill2018quantum, cao2019quantum, zhao2020quantum}. However, to the best of our knowledge\footnote{The statement was correct at the time of writing the paper. However, the proceeding of this workshop will include a paper describing how to improve dynamic software testing using QC~\cite{miranskyy2022using}.}, the question of whether a QC can help software engineers has not yet been explored~\cite{khan2021string, khan2022ep}. 
Thus, we \emph{position} that the SE community may begin exploring the applicability of QC algorithms to SE processes.

With so many algorithms available, how should we start the exploration?
Let us examine eight groups of algorithms: linear equation solvers, differential equation solvers, eigenvalue solvers, data fitters, machine learners, and combinatorial optimizers.

What process did we use to select these algorithms? The selection process\footnote{Conceptually, similar processes are followed in other fields.} consists of two steps: 
\begin{enumerate}
    \item Pick a practical SE use case; 
    \item Investigate the possibility of reducing the use case to an abstract mathematical problem that can be solved efficiently on a QC.
\end{enumerate}
The best candidates for such use cases are those whose solutions show promising results for small SE problems but cannot scale to large ones (due to the high computational complexity of the algorithm solving a given mathematical problem).

Implementing a solution to the use case is relatively straightforward. We can write a program for a classical computer (CC) to convert the data and logic required as an input to a QC algorithm once we have reformulated (reduced) our problem. We can treat many quantum algorithms as black boxes since their implementations are readily available\footnote{ Obviously, if the implementations are unavailable, we will have to implement and test the algorithm ourselves~\cite{miranskyy2019testing, miranskyy2021testing}.}, i.e., we can simply pass the required inputs (via a CC) and observe the outputs~\cite{miranskyy2019testing, miranskyy2021testing}.

\section{Algorithms and their applications}\label{sec:algs}
Table~\ref{tbl:summary} summarizes families of quantum algorithms and gives examples of their potential applications. The details are given below.

\begin{table*}[ht]
\caption{An overview of quantum algorithms and their application to SE use cases.}\label{tbl:summary}
\resizebox{\linewidth}{!}{
\begin{tabular}{@{}llll@{}}
\toprule
Family of algorithms                      & An example of an SE use case          & Potential speed-up  & Applicability horizon \\
                                          & (others are in the text)           & in comparison with CC &  \\
\midrule
Systems of sparse linear equations solvers       & Secure log search                 & Exponential                              & Future                \\
Eigenvalue solver &  Code quality prediction & Exponential                              & Future               \\
Systems of differential equations solvers & Constructing self-adaptive systems & Exponential                              & Future                \\
Data fitting                              & Cost estimation                  & Exponential                              & Future                \\
Machine learning                          & Root cause analysis of failures & Exponential (in some cases)                                  & Future               \\
Combinatorial optimization                & Release planning         & Unknown, but probably modest                                 & Present and Future    \\
Search or string comparison                         & Dynamic testing, log  analysis                      & Quadratic                              & Future                \\
Boolean satisfiability solvers            & False path pruning                & Unknown                                  & Future                \\
\bottomrule
\end{tabular}
}
\end{table*}

\subsection{Systems of sparse linear equations solvers}\label{sec:lin_eq}
Systems of linear equations can be used to solve some SE problems. An example of this is secure search in sensitive log records without exposing their content~\cite{savade2012}.

QCs can be used to solve systems of sparse linear equations. 
The best CC solver runs in $O(N\kappa)$ time, where $N$ is the number of variables and $\kappa$ is the condition number~\cite{harrow2009quantum}.
The seminal QC HHL algorithm proved that QC could approximately solve a sparse system of linear equations in  $O(\log N \kappa^2 / \epsilon)$ time, where $\epsilon$ is the desired precision~\cite{harrow2009quantum}. In other words, this algorithm is exponentially faster than the classical one. 

These results were further improved. Currently, sparse systems of equations can be solved in $\poly(\log N, \log 1/\epsilon)$ time~\cite{childs2017quantum}, where $\poly$ denotes a function upper bounded by a polynomial in its arguments, and dense systems~--- in $O(\sqrt{N} \log N \kappa ^{2})$ time~\cite{wossnig2018quantum}.

\subsection{Eigenvalue solvers}
An eigenvalue solver is used to find the eigenvalue of an eigenvector. Eigenvalues and eigenvectors can be used to assess the importance (centrality) of developers and code modules in order to predict code quality~\cite{pinzger2008can}. They may also be used to identify high-risk software~\cite{DBLP:journals/tse/Neumann02}.

The HHL algorithm, discussed in Section~\ref{sec:lin_eq}, can be used to compute\footnote{Hypothetically, we can also compute the eigenvalues using the quantum phase estimation algorithm~\cite{kitaev1995quantum}. However, its computational complexity is roughly proportional to $O(N^2)$, see~\cite[Section 5.2.1]{nielsen2010quantum} for details.} eigenvalues (leading to exponential speedup over the classical algorithms). The data must be represented as a unitary matrix, but any matrix can be converted to a unitary matrix (at a cost).

\subsection{Systems of differential equations solvers}
Many SE problems can be modelled using systems of linear and non-linear differential equations. They can, for instance, assist in development of self-adaptive software systems~\cite{filieri2015, filieri2017control} and assessing project risks associated with changing requirements~\cite{miranskyy2005modelling,miranskyy2005managing}.

Classically, a system of linear equations with $N$ variables is solved in $O(N)$, whereas QC can solve it in $\tilde O( (\|A\| \Delta t)^2)$ time, where $\|A\|$ is a norm of an $N \times N$ matrix and $\Delta t$ is evolution time~\cite{berry2014high}, which provides an exponential speedup. One more solver, which requires $O(\log N)$ time and is empirically validated, has been proposed~\cite{xin2020quantum}.

QC may also be able to achieve an exponential advantage for non-linear systems~\cite{lloyd2020quantum, Liue2026805118}. For example,  quadratic ordinary differential equations can be solved in $T^2 q \poly(\log T, \log n, \log 1/\epsilon)/\epsilon$, where $T$ is the evolution time, $\epsilon$ is the allowed error, and $q$ measures decay of the solution~\cite{Liue2026805118}.

\subsection{Data fitting}
The least squares method is often helpful for fitting data in various SE areas, such as cost estimation~\cite{962560,DBLP:journals/ese/MittasA10}, software reliability forecasting~\cite{doi:10.1080/16843703.2012.11673290}, and defect prediction~\cite{DBLP:conf/sigsoft/RahmanPHD13, DBLP:conf/icse/Tantithamthavorn16}.

In essence, the least squares method can be reduced to solving a linear equations system~\cite{PhysRevLett.109.050505}. As discussed in Section~\ref{sec:lin_eq},  a QC can solve this problem exponentially faster than a CC.

\subsection{Quantum machine learning}
Machine learning can be used to support most SE activities~\cite{DBLP:journals/sqj/ZhangT03}, from requirements engineering~\cite{DBLP:conf/apsec/IqbalEL18} to maintenance~\cite{DBLP:journals/infsof/AlsolaiR20}.

Several quantum machine learning (QML) algorithms, which may be faster than classical algorithms, have been proposed: supervised and unsupervised machine learning \cite{yom2004advanced, russell2016artificial, p-pqm}, reinforcement learning~\cite{4579244}, and support vector machine~\cite{rebentrost2014quantum}. For an overview of quantum machine learning, refer to~\cite{mishra2021quantum, li2021introduction}. 

While the QML algorithms are faster than their classical counterpart, loading classical data into a QC represents a challenge. This may have negated the efficiency gains from the algorithms themselves~\cite{DBLP:journals/corr/abs-2109-03400}. Recently, Liu et al.~\cite{liu2021rigorous} demonstrated how QC could be used to perform classification using the heuristic quantum kernel method resulting in exponential performance gains over classical algorithms. Thus, QML algorithms can be created for QC to process regular datasets efficiently.

\subsection{Combinatorial optimization}\label{sec:combinatorial_optimization}
Combinatorial optimization techniques, such as integer programming and mixed integer programming (MIP), are essential in SE from release planning and requirements prioritization~\cite{DBLP:journals/isse/SaliuR05, DBLP:journals/infsof/AkkerBDV08, 6824116} to test case prioritization~\cite{DBLP:journals/jss/ZhangYZ014, AHMED2016737} to selecting an optimal set of customers to profile~\cite{DBLP:conf/sigmod/MiranskyyCG09}.

A Quantum Approximate Optimization Algorithm~\cite{farhi2014quantum} and its extensions (e.g.,~\cite{braine2021quantum,gambella2020multiblock,koretsky2021adapting}) have been proposed for gate-based noisy intermediate-scale quantum (NISQ) devices. This is a family of hybrid quantum-classical algorithms in which the optimization is done iteratively on both QC and CC. The algorithms solve the Max-Cut problem, to which mixed binary programming can be reduced~\cite{braine2021quantum}. The computational complexity of these algorithms is not well understood, but empirical evidence suggests they are promising~\cite{crooks2018performance}.

MIP problems can also be solved using adiabatic QCs, such as the D-Wave quantum annealers. They are designed to solve quadratic unconstrained binary optimization (QUBO) problems~\cite{10.1145/2482767.2482797}, and MIP problems can be reformulated as QUBO problems~\cite{8778335, chang2020quantum}.

The nature of quantum annealers makes it challenging to estimate time complexity using the $O$ notation~\cite{mcgeoch2019principles}. Nevertheless, multiple empirical studies, e.g.,~\cite{10.1145/2482767.2482797, dash2013note}, have shown that D-Wave computers might outperform\footnote{Note that a QC cannot guarantee to find an optimal solution to NP-hard problems efficiently~\cite{vazirani2002survey}.} top-of-the-line heuristic solvers running on a CC. Moreover, the latest D-Wave machines can solve optimization problems with up to one million variables (in hybrid mode, i.e., by also leveraging CC resources)~\cite{DWaves5063:online}. Therefore, practitioners could benefit from them right now.

In the future, gate-based QCs may prove useful for solving these types of problems (although the increase in speed is likely to be modest when compared to CC~\cite{davis2021cutting}), while quantum annealer technology is already on the market.

\subsection{Search or String comparison}
String comparison and pattern matching are omnipresent in SE, appearing in areas of static and dynamic analysis, such as test selection and generation~\cite{malaiya1995antirandom,wu2008antirandom,mrozek2012antirandom,hridoy2015regression}, code coverage inspection~\cite{hridoy2015regression,horvath2019code}, log or trace analysis~\cite{du2004anomaly,miranskyy2007iterative,miranskyy2008sift,miranskyy2016operational}, and cybersecurity~\cite{du2004anomaly,taheri2020similarity,kanuparthi2016controlling,dario2017detecting}. The authors of these papers represent a collection of SE artifacts as symbols in a string.

QC may speed up string comparison and pattern matching. Given a pattern of length $M$ and a string of length $N$, on a CC finding the occurrence of pattern in the string takes $O(M+N)$ operations, while QC may reduce it to $O(\sqrt{M}+\sqrt{N})$~\cite{DBLP:journals/jda/HariharanV03} or even to $\tilde O(\sqrt{N})$~\cite{niroula2021quantum}. These QC algorithms are extensions of the Grover's algorithm~\cite{grover1996fast}, hence the quadratic speed up. 

Patterns and data may be $d$-dimensional. On a CC finding an occurrence of such a pattern requires $\tilde\Omega(N^{d/2} + (N/M)^d)$, while on a QC it takes only $\tilde O\left((N/M)^{d/2}2^{O\left(d^{3/2}\sqrt{\log{M}}\right)}\right)$~\cite{DBLP:journals/algorithmica/Montanaro17}.
Thus, QC may be helpful in SE use cases that can be reduced to string comparison. 

Finally, as mentioned in Section~\ref{sec:intro}, this workshop proceedings include a paper that explores how to speed up exhaustive and non-exhaustive dynamic testing of programs written for a CC by combining quantum counting~\cite{boyer1998tight, brassard1998quantum, aaronson2020quantum} and Grover's search algorithms~\cite{miranskyy2022using}. On the CC, the computational complexity of these techniques is $O(N)$, where $N$ represents the count of combinations of input parameter values passed to the software under test. On the QC, the complexity is reduced to $O(\varepsilon^{-1} \sqrt{N/K})$, where $K$  denotes the number of inputs causing errors and $\varepsilon$ is a relative error of measuring $K$.

\subsection{Boolean satisfiability solvers}
Many SE use cases (e.g., static analysis software model checking~\cite{cimatti2012software}, false path pruning~\cite{chelf2007sat}, and test suite reduction~\cite{arito2012application}) can be formulated as a Boolean satisfiability (SAT) problem with a large number of literals $k$.

QCs show promise at solving SAT problems. Adiabatic QCs can be used for solving such problems~\cite{bian2020solving}. However, as was discussed in Section~\ref{sec:combinatorial_optimization}, it is unclear what the time complexity of this approach is.

As for the gate-based QCs, currently, the QC algorithm (based on Grover's search)~\cite{cerf2000nested, Qiskit-Textbook} for a $k$-SAT problem with $k=3$ is faring worse than the top algorithm for a CC~\cite{hansen2019faster}: $O(1.414^n)$ versus $O(1.307^n)$, where $n$ is the number of variables. However, the unproven but popular Strong Exponential Time Hypothesis~\cite{hansen2019faster} conjectures that the computational complexity of $k$-SAT, given by $O(c^n)$, converges to $O(2^n)$ as $k \to \infty$ when executed on a CC. Thus, a QC-based SAT solver may outperform a CC-based one for a large value of $k$~\cite{cerf2000nested, Qiskit-Textbook}.
Of course, $k$-SAT may be reduced to 3-SAT in polynomial time, but there is an overhead. 
Therefore, it may be advantageous to use QC for SE use cases that can be converted to $k$-SAT (when $k \gg 3$). 

\section{Applicability horizon}
In Section~\ref{sec:algs}, we examined eight groups of quantum algorithms. When will software engineers be able to benefit from them? QC is not required for such problems. As mentioned in Section~\ref{sec:intro}, QC is most useful for large problems. Thus, the question can be rephrased as, when will we see a QC capable of handling large inputs and solving complex problems efficiently? 

Those using combinatorial optimization can now take advantage of D-Wave's adiabatic solvers. They are available as a Cloud service directly from the manufacturer~\cite{leap:online} or through third-party providers, such as Amazon Web Services Braket~\cite{braket:online}. 

The remaining algorithms require gate-based NISQ devices. Let us look at specifics.

Linear and eigenvalue solvers, as well as least squares data fitting, utilize the HHL algorithm, which requires $\propto \log N$ qubits, where $N$ is the number of variables and $\propto$ denotes ``proportional to''. The differential equation solvers also encode the variables using  $\propto \log N$ qubits~\cite{Liue2026805118}. Thus, a NISQ device with a few dozen qubits may be able to solve huge computational problems (assuming that the problem can be formulated using a sparse system of equations). NISQ devices now have up to 127 qubits~\cite{IBMUnvei91:online}, but the level of noise in these devices prevents them from being used to solve practical problems. However, this is an engineering problem that will hopefully be resolved within the next five to ten years~\cite{preskill2018quantum}.

Quantum machine learning requires $\propto \log(vw)$ to $\propto vw$ qubits, where $v$ is the number of features and $w$ is the number of observations~\cite{schuld2018supervised,lloyd2020quantum}. Moreover, the iterative machine learning approaches may partition the inputs~\cite{perez2020data}, enabling efficient computations using a few qubits. Accordingly, some machine learning algorithms may become practical in five to ten years, similar to HHL-based algorithms\footnote{HHL is leveraged in quantum machine learning too~\cite{duan2020survey}.}.

At the moment, SAT solvers, combinatorial optimization  (on a NISQ device), and string comparison\footnote{Furthermore, Grover's algorithm and its descendants require at least $O(n)$ time to prepare the database. It is unclear whether a basic string comparison on a QC will be useful until this problem is solved.} require $\propto x$ qubits, where $x$ is the number of variables passed to the solver or optimizer or the number of characters in the string. Thus, it will take longer for these algorithms to become practical (especially because the computational speedup of these algorithms is modest). In theory~\cite{bian2020solving}, adiabatic solvers may speed up SAT solvers (especially in hybrid mode). However, more empirical evidence is needed to support this.

Finally, note that NISQ devices may outperform CCs for specific tasks (especially if the level of noise is reduced), but fault-tolerant QCs (FTQCs)~\cite{nielsen2010quantum} will likely be required for most applications. FTQCs will employ error-correcting schemes that will enable us to operate on logical/ideal qubits (constructed from groups of physical qubits)~\cite{chao2018fault}. FTQCs are not yet available, but small ones may appear within ten years~\cite{sevilla2020forecasting}. For the problems discussed here, quantum algorithms executed on FTQCs are guaranteed to outperform classical algorithms on CC. 

\section{Conclusions}\label{sec:conclusion}
We explored eight families of quantum algorithms that may enable SE techniques that do not scale well. These techniques appear throughout the software development lifecycle (from requirements engineering to maintenance).

Except for Mixed Integer Programming (which may be used in, e.g., requirements or test cases prioritization), such algorithms will not be practical for at least five years (i.e., when QCs become sufficiently powerful). However, the community can develop the algorithms and software foundation in time for the hardware to arrive. 

\section*{Acknowledgement}
We thank anonymous reviewers for their thoughtful comments and suggestions!

We also would like to acknowledge Canada's National Design Network for facilitating this research, specifically through their member access to the IBM Quantum Hub at Institut quantique.

This research is funded in part by NSERC Discovery Grants No. RGPIN-2015-06075 and RGPIN-2022-03886.

\bibliography{reference}
\end{document}